# Integrated Photonic Encoder for Terapixel Image Processing

Xiao Wang [1, +], Brandon Redding [2, +], Nicholas Karl [3], Christopher Long [3], Zheyuan Zhu [4], Shuo Pang [4], David Brady[1], Raktim Sarma[3, 5, *]

[+] Equal contribution

[*] rsarma@sandia.gov

[1] Wyant College of Optical Sciences, University of Arizona, Tucson, Arizona, USA

[2] U.S. Naval Research Laboratory, Washington, DC, USA

[3] Sandia National Laboratories, Albuquerque, New Mexico, USA

[4] CREOL, The College of Optics and Photonics, University of Central Floria, Orlando, Florida, USA

[5] Center for Integrated Nanotechnologies, Sandia National Laboratories, Albuquerque, New Mexico, USA

## Abstract (150 words)

Modern lens designs are capable of resolving >10 gigapixels, while advances in camera frame-rate and hyperspectral imaging have made Terapixel/s data acquisition a real possibility. The main bottlenecks preventing such high data-rate systems are power consumption and data storage. In this work, we show that analog photonic encoders could address this challenge, enabling high-speed image compression using orders-of-magnitude lower power than digital electronics. Our approach relies on a silicon-photonics front-end to compress raw image data, foregoing energy-intensive image conditioning and reducing data storage requirements. The compression scheme uses a passive disordered photonic structure to perform kernel-type random projections of the raw image data with minimal power consumption and low latency. A back-end neural network can then reconstruct the original images with structural similarity exceeding 90%. This scheme has the potential to process Terapixel/s data streams using less than 100 fJ/pixel, providing a path to ultra-high-resolution data and image acquisition systems.

## Introduction

From the invention of photography until the early 1990's the function of a camera was to record analog images. With the development of solid-state focal planes and digital coding, however, this function changed, and modern digital cameras act as transceivers that transform massively parallel optical data streams into serial coded electronic data that is processed 1 pixel at a time. While this paradigm shift introduced numerous advantages, the power consumption associated with electronic digital processing is now the bottleneck limiting image data acquisition rates. In contrast, compact lens designs are capable of resolving >10 gigapixels of transverse resolution [1, 2], while advances in multimodal imaging systems capable of acquiring spectral, polarization, temporal, and range information could enable future imaging systems to acquire (Tera)$10^{12}$ pixels per second of data. However, the power consumption and resulting coding capacity of opto-electronic transceivers is the primary barrier to achieving such systems.

In current digital electronics-based image processing systems, electrical power is proportional to the number of mathematical operations performed on each pixel. Conventional image signal processing (ISP) systems perform 100-1000 operations per pixel to first condition (e.g. compensate for pixel non-uniformity, hot-pixels, denoising etc.) and then compress the image data stream, resulting in a per pixel cost of 0.1-1 microjoule [3]. In addition, while image compression is required for most remote sensing applications, many of these pixel conditioning operations are performed at the front end regardless of whether they are necessary for a given application. Recently, different ISP approaches such as blind sensor head compression were proposed to reduce the number of operations per pixel [3]. Blind compression, in this context, can be understood as implementing the first layer or two of a deep neural network-based auto-encoder on the read-out data stream. While this approach substantially reduces the number of operations per pixel, the per pixel power cost still remains unacceptably high, at ~ 0.01-0.1 microjoule [3].

Future Terapixel/s imaging systems will require an alternative ISP approach with dramatically higher throughput and lower power consumption. In this work, we explore the potential for analog photonics to perform parallel pixel processing with high-speed and low energy consumption. Our approach builds on the neural network framework proposed in the blind compression work [3], which allows the key front-end image processing task to be accomplished using a single matrix vector multiplication. Fortunately, analog optical computing is particularly well-suited for this type of operation and is being explored for a variety of matrix-multiplication-heavy computing applications [4-6]. The key advantage is that optical computing engines can perform matrix multiplication with energy consumption that scales linearly with the dimension of the input dataset ($N$) as opposed to the quadratic scaling ($N^2$) inherent to electronic approaches. In addition, optical computing engines are able to process $N$ pixels in parallel with an overall speed that can exceed 100 GHz, only limited by the speed of the optical modulators and photodetectors used to encode the input data and record the result [7-9].

In this work, we show that an optical image processing engine can take advantage of these unique features to enable high-speed image compression with the potential for orders-of-magnitude lower power consumption than current techniques. Our approach is based on a passive, CMOS-compatible silicon photonics device that performs the matrix-vector multiplication required for front-end image processing. We experimentally demonstrate image compression with a ratio of 4:1 and develop a back-end neural network capable of reconstructing the original images with an average peak signal-to-noise ratio (PSNR) ~ 25 dB and structural similarity index measure (SSIM) ~ 0.9, comparable to common electronic software-based lossy compression schemes such as JPEG [10]. Analysis of the throughput and power consumption for our optical image processing engine indicates that this technique has the potential to encode terapixel per second data streams utilizing <100 femtojoules per pixel—representing a >1000x reduction in power consumption compared with the state-of-the-art electronic approaches. This improvement in throughput and power consumption is made possible by parallelization of pixel processing and transferring the majority of the image processing and conditioning tasks (e.g., denoising, linearization, etc.) from the front-end encoding interface to the back-end decoding interface. This trade-off is particularly attractive for remote sensing and imaging applications in which image reconstruction is only performed on a "need to know" basis and is usually conducted at a remote cloud site with better access to power. Finally, by constructing the optical image compression engine on a silicon photonics platform, this approach meets the size-weight-and-power and integration requirements for a wide range of surveillance and remote sensing applications.

## Results

Operating principle of the image encoder:

The optical image compression scheme relies on an auto-encoder neural network framework in which the compressed image is naturally formed at the "bottleneck" layer in a neural network [11]. This approach has gained traction in recent years due to its ability to simultaneously perform data compression, dimensionality reduction, and denoising [11-13]. In our implementation, the first half of the neural network (mapping the original image data to the compressed image at the bottleneck layer) is performed optically while the second half of the network (reconstructing the image) is performed using digital electronics. A schematic of the optical encoder and corresponding neural network structure is shown in Figure 1. Our hybrid opto-electronic auto-encoder takes advantage of the fact that most neural encoding networks are not very sensitive to details of the feature map (i.e., the weights and connections) implemented by the first few layers. In fact, researchers have shown that the first few layers can often be assigned random weights without compromising performance [14]. This allows us to use a pre-designed, passive photonic device to perform the transform used in the first layer of the auto-encoder network. In this case, we designed the photonic layer to perform local kernel-like random transforms on small blocks of the image at a time. This random encoding scheme was selected based on compressive measurement theory, which has shown that random transforms are ideal for a variety of dimensionality reduction compression tasks [15-19].

As shown in Figure 1, the silicon photonics-based image encoder consists of $N$ single mode input waveguides, each with a dedicated modulator, followed by a multimode waveguide region, a random encoding layer, and $M$ photodetectors. A laser (not shown) provides light with equal amplitude to the $N$ input waveguides where the $N$ modulators encode a $\sqrt{N} \times \sqrt{N}$ pixel block of the input image onto the amplitude of light transmitted through each waveguide. Light from each waveguide is then coupled into a multimode waveguide region before scattering through the random encoding layer and finally reaching the photodetectors. The random encoding layer consists of a series of randomly positioned scattering centers fabricated by etching air holes in the silicon waveguiding layer (see the Methods section for more details). Since the optical device operates in the linear regime, we can describe the encoding process using a single transmission matrix (*T*), which relates the input (*I*) to the transmitted output (*O*) as *O=TI*, where *I* is a $N \times 1$ vector, *O* is a $M \times 1$ vector, and *T* is a $M \times N$ matrix. By forcing *M* to be less than *N*, the device effectively performs a single matrix multiplication in order to compress an *N* pixel block of the original image into *M* output pixels. Since the random encoding layer is entirely passive, this compression process can be extremely fast, operating on *N* pixels in parallel at speeds limited only by the modulators and photodetectors. In addition, the energy consumption scales linearly with *N* (i.e. the number of modulators), even though the device performs $M \times N$ operations.

The local kernel size, *N*, is a key parameter driving the performance of the photonic image processing engine. While using smaller kernel-like transforms reduces the data throughput, since the device can only compress *N* pixels at a time, smaller kernels also have several advantages. First, local transforms maintain the spatial structure of the original image, which tends to improve the image reconstruction, as we discuss in the next section. Second, the kernel approach can be used to compress arbitrarily large images without requiring a corresponding increase in the numbers of modulators and detectors. Third, using these local transforms helps to isolate noise from a given pixel (e.g. a hot pixel), which could otherwise spread across the entire compressed image. Finally, since this compression scheme effectively maps the input image blocks to speckle patterns, using a large kernel could lead to low contrast speckle which can also degrade the image reconstruction similar to the trend observed in speckle-based spectrometers [20].

Effect of kernel size and kernel type on image compressibility:

To determine the effect of kernel size on image compression, we performed numerical simulations of the image compression and reconstruction process using images taken from the DIV2K dataset [21] and synthetically generated random *T* matrices. To reduce the computation time, the images were converted to grayscale and cropped to a resolution of $512 \times 512$ pixels. The dataset consisted of 4152 images divided into a 3650-image training dataset and a 502-image validation dataset. In this case, the compression process was simulated by multiplying each $\sqrt{N} \times \sqrt{N}$ block of an image by a numerically generated random matrix consisting of real, positive numbers uniformly distributed between 0 and 1. We then trained a neural network to reconstruct the original image from the compressed image (see Methods for a detailed description of the neural network architecture and training routine). Finally, we used the test images to evaluate the reconstructed image fidelity after compression using kernels of varying sizes.

An example image from the set of test images is shown in Fig. 2(a) along with the compressed images obtained using 8×8 (Fig. 2b) and 32×32-pixel kernels (Fig. 2e). In this case, we fixed the compression ratio ($N:M$) at 8:1 and the images were compressed from 512×512 to $8 \times (64 \times 64)$ or $128 \times (16 \times 16)$ pixel datacubes. The reconstructed images using the two kernel sizes are shown in Fig. 2(c, f). Using a smaller kernel size clearly retains more of the spatial structure in the compressed image (Fig. 2(b)) resulting in a higher fidelity reconstruction. The average peak signal-to-noise ratio (PSNR) and the structural similarity index measure (SSIM) of the reconstructed images in the test image dataset are shown in Fig. 2(d) as a function of kernel size. We found that smaller kernels generally result in higher quality image reconstruction. Unlike purely random data—which can be efficiently compressed using large random matrices—image data naturally includes spatial structure which is important to maintain. In general, the optimal kernel size will depend on the type of images being compressed and will be impacted by factors such as the sparsity and spatial frequency content of the images. For benchmarking, we compared our compression technique using an 8×8 kernel to standard JPEG compression of ten randomly chosen images from the same test dataset and with the same 8x compression factor. We found that the JPEG compression provided a PSNR (SSIM) of 38.82 ± 3.12 (0.95± 0.02), comparable to our encoding scheme which achieved a PSNR (SSIM) of 34.45 ± 3.39 (0.93 ± 0.03).

In addition to kernel size, we also investigated two types of random kernels. In Figure 2, we presented simulations using synthesized random $T$ matrices that were real and positive. This simulated the compression process if light coupled to each of the input waveguides was effectively incoherent. For example, a frequency comb or other multi-wavelength source could be used to couple light at different wavelengths into each waveguide (to be more quantitative, light in each waveguide should be separated in frequency by at least ~10× the detector bandwidth to minimize interference effects) [22]. In this case, the speckle patterns formed by light from each waveguide would sum incoherently on the detectors and the compression process can be modeled using a random $T$ matrix that is real-valued and non-negative. The second case we considered is using a complex-valued field $T$ matrix, in which each element in $T$ was assigned a random amplitude and phase. In this case, the compressed image was obtained as the square-law detector response: $O = (T\sqrt{I})(T\sqrt{I})^*$. This case simulates the effect of coupling a single, coherent laser to all the input waveguides at once such that the measured speckle pattern is formed by interference between light from each waveguide.

To evaluate the trade-off between real and complex transforms, we evaluated the reconstructed image quality at varying noise levels. In general, noise could be introduced in the original image formation process (e.g. due to low-light levels or imperfections in the imaging optics), through the camera opto-electronic conversion process (e.g. due to pixel non-linearity or the limited bit depth of the camera pixels), or through the optical compression process described in this work (e.g. due to laser intensity noise, environmental variations in the $T$ matrix, or simply shot noise at the detection stage). To simulate the effect of noise on the reconstruction of the compressed images, we numerically added gaussian white noise to the compressed images. Figure 3(a, d) shows the same test image evaluated in Fig. 2, compressed using either a real-or complex-valued 8x8 pixel $T$ matrix. In this case, gaussian white noise with an amplitude equal to 2% of the average signal level in the image (corresponding to an SNR of 50) was added to each compressed image. The reconstructed images using the real and complex $T$ matrices are shown in Fig. 3(b, e). At 2 % noise level (SNR = 50), the reconstructed images are only marginally worse than the reconstructed image without noise shown in Fig. 2(c) (PSNR = 25.1 dB with noise vs PSNR =26.9 dB without noise for the case of a real transform). This measurement confirms that the autoencoder framework is relatively resilient to noise, which is consistent with prior applications of autoencoders for denoising tasks. This resilience could also enable the system to forego the energy-intensive image conditioning by encoding raw image data and relying on the back-end neural network to compensate for noise due to effects such as pixel non-uniformity. In Fig. 3(c, f), we present the average PSNR and SSIM for reconstructed test images that were compressed using either real valued or complex valued $T$ matrices, as a function of the SNR of the compressed images. These simulations showed that at relatively high SNR (>50), the real and complex valued $T$ matrices provided comparable performance. However, at lower SNR, the complex valued $T$ matrices provided more robust image compression due to the higher contrast in the compressed images obtained using a complex $T$ matrix.

Experimental image compression and denoising:

To experimentally validate our image processing approach, we fabricated a prototype device on a silicon photonics platform. The experimental device included $N=16$ single mode input waveguides connected to the scattering layer through a multimode waveguide. The multimode waveguide region allowed light from each single mode waveguide to spread out along the transverse axis before reaching the random scattering layer. This ensured that we obtained a uniformly distributed random transmission matrix without requiring an excessively long random scattering medium, which would introduce excess loss through out-of-plane scattering. To illustrate the impact of the multimode waveguide, we performed full-wave numerical simulations of single mode waveguides either connected directly to the scattering layer or connected through an intermediate multimode waveguide region. In the first case, shown in Fig. 4(a), the scattering layer was not thick enough for light to fully diffuse along the transverse axis, resulting in a high concentration of transmitted light near the position of the input waveguide. In terms of a $T$ matrix, this resulted in stronger coefficients along the diagonal, rather than the desired, uniformly distributed random matrix. We then simulated the effect of adding a 32 $\mu m$ multimode waveguide between the single mode input waveguide and the scattering layer. As shown in Fig. 4(b), the multimode waveguide allowed light from the single mode waveguide to extend across the scattering layer, resulting in a transmitted speckle pattern that was uniformly distributed.

The device was fabricated using a standard silicon-on-insulator wafer with a 250 nm thick silicon layer. The fabricated device consisted of 16 single mode input waveguides connecting the device to the edge of the chip. The waveguides were 450 nm wide and separated by 3 $\mu m$ (corresponding to a spacing of ~ 2$\lambda$ at a wavelength of 1550 nm to minimize evanescent coupling). All 16 waveguides were connected to a 55.2 $\mu m$ wide, 120 $\mu m$ long multimode waveguide region, followed by a 30 $\mu m$ long scattering region. The scattering region consisted of randomly placed 50 nm radius cylinders with a 3% filling fraction etched in the silicon waveguiding layer. The scattering layer parameters were empirically optimized to achieve a transmission of ~ 20% [23, 24]. To minimize leakage of light at the edges of the scattering layer, we added a full band-gap photonic crystal layer on the sides of the scattering layer [23-25]. Since this initial prototype did not include integrated photodetectors, we etched a ridge in the silicon waveguiding layer after the scattering region. This allowed us to record the light scattered out-of-plane from this ridge to measure the optical power which would be recorded if detectors were integrated in the device. Scanning electron microscope images of the fabricated device are shown in Fig. 4(c).

To test the device, we first measured the $T$ matrix by coupling an input laser operating at a wavelength of 1550 nm into each single mode waveguide and recording the speckle pattern scattered from the detection ridge after the scattering layer using an optical microscope setup. A typical image recorded using the optical setup is shown in Fig. 4(d).

In order to account for experimental noise in the image compression and recovery process, we recorded two sequential $T$ matrices, as shown in Fig. 5(a, b). The $T$ matrix was highly repeatable, as revealed in Fig. 5(c), which shows the difference between the two measurements. A histogram of the difference in the matrices, shown in Fig. 5(d), indicates a gaussian-like random noise with amplitude ~1% of the average signal value (corresponding to a measurement SNR~100). As shown in Fig. 3, at this SNR, both real and complex transformations provide similar results in terms of image reconstruction. This implies that we can use the experimentally measured intensity transmission matrix for image compression.

To convert the raw measured transmission matrix into the $T$ matrix used for compression, we selected 4 non-overlapping spatial regions along the "output" ridge shown in Fig. 4(d). This corresponds to selecting 4 columns of the matrix shown in Fig. 5(a). This updated $T$ matrix had dimensions of $16 \times 4$, providing a compression factor of 4. We then used this experimental matrix to train the back-end neural network required to reconstruct the original image. Note that we included noise in the training process by adding gaussian noise with the same 1% variance measured experimentally. Finally, we compressed the test images in the DIV2K dataset while again adding random noise with 1% variance. A typical compressed image using the experimentally measured $T$ matrix is shown in Fig. 5(e) and the corresponding reconstructed image is shown in Fig. 5(f). Excellent agreement between the original and reconstructed images can be seen with PSNR = 26.02 dB and SSIM = 0.91. We repeated this process for the entire set of test images and obtained an average PSNR of $26 \pm 4$ dB and SSIM of $0.9 \pm 0.07$. Additional examples of compressed and reconstructed images and histograms of PSNR and SSIM of the reconstructed images are shown in the Supplementary Information.

Finally, in addition to showing that this approach is robust to noise introduced in the analog photonic image compression step, this demonstration also illustrates how this technique could be used for image denoising. From the perspective of the back-end image reconstruction neural network, noise added during the original image acquisition process (e.g., due to pixel noise, non-uniform responsivity, or simply low light levels) is equivalent to noise added during the image compression step (as tested explicitly here). Thus, this work also highlights the potential for this technique to move the energy-intensive image conditioning and denoising steps to the back-end image reconstruction stage.

Predicted energy consumption and operating speed for the photonic image processor:

As described above, our encoding and compression technique can be reduced to a matrix multiplication operation. In order to compare the power consumption using our photonic approach with a traditional electronic scheme, we estimated the energy per multiply-accumulate (MAC) operation using both approaches. Electronic hardware accelerators have been thoroughly optimized to reduce the power consumption per MAC. Table 1 in the Supplementary Information summarizes the energy consumption typical of mainstream electronic architectures, with the most efficient schemes reaching ~1 pJ/MAC.

The power consumed by the photonic image processing engine includes contributions from the power consumed by the laser, the optical modulators, and the photodetectors. To estimate the required laser power, we first estimated the required detected power to provide sufficient signal-to-noise for accurate image compression. Assuming shot-noise limited detection, we can express the required optical power reaching each photodetector as [26]:

$$P_{Rx} = 2^{2ENOB} q f_0 / \mathcal{R} \qquad (1)$$

where *ENOB* is the required effective number of bits, $q$ is the charge of a single electron (1.6 x 10$^{-19}$ coulombs), $f_0$, is the operating frequency of the modulator (and the detector baud rate), and $\mathcal{R}$ is the responsivity of the photodetector in units of $A/W$. The *ENOB* can be related to the measurement SNR in dB as $SNR = 6.02 \times ENOB + 1.72$ [26]. In the energy consumption calculations below, we assumed a required *enob* of 6, corresponding to a measurement SNR of 38 dB, which provides significant margin compared with the experimentally measured SNR of 17 dB. Based on the required power at the detector, we can work backwards to estimate the required laser power as

$$P_{laser} = \frac{N \times P_{Rx}}{T_{mod} T_{scatter}} \quad (2)$$

where $N$ is the number of pixels in an image block, $T_{mod}$ is the transmission through the optical modulators, and $T_{scatter}$ is the transmission through the scattering medium. The electrical power required to drive the laser can then be written as $P_{laser}/\eta$, where $\eta$ is the wall-plug efficiency of the laser. The factor of $N$ in equation 2 implies that the multimode waveguide and scattering region support $N$ spatial modes (the minimum required to efficiently couple light from $N$ single-mode input waveguides) and each detector collects, on average, $1/N$ of the light transmitted through the scattering medium. In our preliminary experiment, presented above, a slightly larger multimode waveguide than required was used to simplify the experiment. As a result, the optical power was distributed over more than $N$ modes in our initial demonstration. In the future, adiabatically coupling the single-mode input waveguides into an $N$-mode multimode waveguide would optimize the power efficiency.

The power required by the optical modulators can be expressed as [27]

$$P_{Mod} = \frac{1}{2} C_{Mod} V_{pp}^2 f_0 \quad (3)$$

where $C_{Mod}$ is the capacitance and $V_{pp}$ is the peak-to-peak driving voltage of the modulator. The power required by the photodetectors can be approximated as

$$P_{PD} \approx V_{bias} \mathcal{R} P_0 \quad (4)$$

where $V_{bias}$ is the bias voltage of PN junction. The total electrical power consumed by the photonic image processing engine can then be calculated as

$$P_{total} = P_{laser}/\eta + N \times P_{mod} + M \times P_{PD} \quad (5)$$

Since the total number of MACs per second is $N \times M \times f_0$, the energy consumption per MAC is given by $P_{total} / (N \times M \times f_0)$. After substituting equation 1 into the expressions for $P_{laser}$ (Eq. 2) and $P_{PD}$ (Eq. 4), we see that the total energy consumption per MAC is independent of the modulation frequency.

To quantitatively compare the energy per MAC required by an optimized photonic processing engine with a conventional electronic GPU, we assumed typical specifications for the opto-electronic components. $C_{Mod}$ is usually on the order of 1fF, $V_{pp}$ is ~ 1V, $V_{bias}$ is typically 3.3V, and $\mathcal{R}$ is typically ~1mA/mW at a wavelength of 1550 nm [28, 29]. In addition, typical insertion loss for high-speed optical modulators is ~ 6.4 dB ($T_{mod} = 0.27$) and the wall-plug efficiency for distributed feedback lasers is $\eta = 0.2$ [30, 31]. The transmission through the experimental scattering medium is $T_{scatter} = 0.2$. The estimated energy consumption per MAC as a function of the image block size $N$ is shown in Figure 6. The energy required by the photonic image processor decreases rapidly with image block size. We also find that the majority of the power consumption is driven by the laser. As a result, lower power consumption could be achieved if a lower $ENOB$ was sufficient for a given application, which would enable a lower power laser (see Fig. 3(c, f) for an analysis of the trade-off between image reconstruction quality and the SNR of the compressed image). Nonetheless, we find that for an image block size of $8 \times 8$ pixels, the photonic image processor has the potential to provide 100x lower power consumption than a typical GPU. Although the photonic processor is even more efficient when using larger image blocks; this can degrade the image reconstruction, as shown in Fig. 2(d). In the future, alternative inverse-designed transforms could enable large pixel blocks without sacrificing image reconstruction fidelity.

We can also use this framework to estimate the energy consumption per pixel, which is calculated as $P_{total}/(N \times f_0)$. The energy per pixel is independent of both modulation frequency and the size of the pixel blocks, and for an $ENOB$ of 6 and the parameters listed above, can be as low as 72 fJ. This is dramatically lower than the ~ 0.1 $\mu$J used in existing image processing systems; however, the latter also includes the power required to operate the pixels and the analog-to-digital conversion process used to extract the signal recorded by each pixel. Nevertheless, since more than 50% of the energy consumed by standard electronic image processing systems is dedicated to image compression and conditioning, our opto-electronic approach can contribute significantly to reducing the overall energy consumption.

Finally, the device throughput, in terms of pixels/second, can be estimated as $N \times f_0$. Assuming an image block size of $8 \times 8$ ($N = 64$), this approach can process a Terapixel/s of image data using a clock speed of ~16 GHz, which is readily achieved using standard optical modulators and photodetectors [7-9].

## Discussion

In summary, we proposed a CMOS-compatible silicon photonics approach for large scale image processing. Our approach performs image compression and de-noising using an auto-encoder framework in which the first layer of the network is performed using analog photonics, while the back-end image reconstruction is performed using digital electronics. We showed, through simulations and proof-of-concept experiments, that this approach enables image compression with comparable quality to standard compression techniques such as JPEG. In addition, by processing blocks of pixels in parallel and leveraging the efficiencies of analog photonics, our scheme has the potential to process Terapixels/s of image data with energy consumption that is orders of magnitude lower than electronic approaches (e.g. 10 fJ/MAC using an 8x8 pixel block). In contrast to the prevailing image processing approach, which performs a large number of image conditioning tasks at the front end, our approach is designed to compress the raw image data and use the neural network back-end to both reconstruct the original image and perform the image denoising and conditioning operations. In this work, we presented a passive, proof-of-concept device and showed experimental compression using a real-valued transmission matrix. By training the back-end digital neural network using the experimentally calibrated T matrix, we show that this approach is extremely robust to noise introduced from different sources including fabrication imperfections. In the future, integration of active components (modulators and detectors) will enable operation using a complex-valued transmission matrix (see Supplementary Information for more details), which our simulations predict will be even more robust to noise. Although this work focused on compression and denoising of grayscale images, the general approach could be used to compress red-green-blue (RGB) images, hyperspectral, or time-series image data. Finally, this general scheme is amenable to a variety of imaging processing tasks other than compression including inference or classification. Again, the analog photonic transform could form the first layer of a neural network, accelerating the initial time and energy-intensive processing of high-dimensional image data, while relying on back-end digital electronics to complete the network. By tailoring this back-end neural network, the same photonic image processing engine could be applied to a variety of image processing and remote sensing applications.

## Methods

<u>Neural processing algorithm at the backend:</u> All the images tested were from the DIV2K dataset. A dataset of 4152 grayscale images was generated, each with a resolution of 512×512 pixels, through cropping and grayscale conversion. The dataset was divided into a training set of 3650 images and a validation set of 502 images. To study the image compression process, numerically generated random transmission matrices as well as experimentally measured transmission matrices were utilized as the encoding matrix. By using the original images as the ground truth and the compressed measurements as input, a convolutional neural network (CNN) was successfully trained to establish a correlation between the ground truth and the compressed images. The neural networks used were constructed based on Deep ResUnet [32] and ResUNet++ [33]. In order to investigate the impact of different compressive kernel sizes on the networks' ability to reconstruct compressed images, four different kernel sizes were explored (Figure 2). The network architecture for the 4×4 kernel size consisted of 1 initial layer, 11 residual blocks, 2 down sampling layers, 4 up sampling layers, and 1 final convolutional layer. The residual block architecture was based on the residual neural network [34], and each block consisted of a Conv(3×3)-BN-LeakyReLU-Conv(3×3)- BN-[Conv(1×1)]-LeakyReLU block, with a Conv(1×1) layer added in the residual connection (marked with brackets). The downsampling (upsampling) rate in each downsampling (upsampling) layer was 2. The downsampling layer comprised of a Conv(3×3)-BN block, where the convolutional layer had padding =1 and stride = 2. The upsampling layer used ConvTranspose2d with kernel size 2×2 and stride=2. The initial layer was a Conv(3×3)-BN-LeakyReLU-Conv(3×3)-Conv(3×3) layer, which produced 64 feature maps, and the final convolutional layer had kernel size 1×1. For the networks with other kernel sizes, we added additional upsampling layers and residual blocks to maintain the number of layers and the size of the final output tensor. The network was trained with the mean squared error loss, Xavier initialization [35], Adam optimizer [36] with learning decay rate 0.1 per 400 epochs, and initial learning rate 0.001 for 800 epochs in PyTorch. The training was performed on four Nvidia V100S GPUs with a batch size of 64.

<u>Full-wave electromagnetic simulations:</u> The simulated fields shown in Figure 4 (a, b) were obtained using the finite element method, (COMSOL with the Electromagnetic Waves, Frequency Domain interface) in a 2D geometry. The length of the simulated devices was scaled down to reduce the computational time and consisted of a scattering region 15 µm in length followed by a 10 µm long output port. Perfect conductors were substituted for the photonic crystal reflectors on the top and bottom sides of the scattering region. The scattering region contained a randomly generated set of 6180 holes with radius = 50 nm. Perfectly matched layer (PML) boundary conditions were used on the boundaries of the observation region. Ports were used to excite the input waveguides (width 0.45 um) which had a pitch of 3.45 um with 16 total inputs in all giving a total device width of 55.2 µm (similar to fabricated structure). The pre-scattering-region length corresponding to multimode waveguide region between the single mode waveguide inputs and the scattering region was varied between 50 nm and 32 µm and had a 4 µm air buffer on the side edges before absorbing PML boundaries. The effective material index of the silicon domains that we used for the 2D simulations was determined from 3D simulations and it was found to be $n_e = 2.83$. The effective material index of the air domains was $n_e = 1$. For the simulation, a triangular mesh was used with 1,256,090 elements with an average element quality of 0.92. A boundary mode analysis step was performed for each input waveguide port and the model was simulated at $\lambda = 1550$ nm. For analyzing the electromagnetic fields in the output region, the fields were exported on a regular grid with 100 nm steps in *X* and *Y*.

Sample fabrication: The silicon-photonics image encoder was fabricated using commercially available silicon-on-insulator (SOI) wafers. The wafers consisted of 250 nm silicon on top of a 3 µm buried oxide. The encoder was fabricated using a positive tone ZEP resist followed by electron beam lithography and inductively-coupled plasma reactive ion etching. The encoder consisted of $N = 16$ input single mode waveguides and width of each waveguide was 450 nm. At the output, a ridge was fabricated to scatter the light out-of-plane which was then measured to determine the transmission matrix of the encoder. The input waveguides were separated by 3.45 µm to minimize cross-coupling. The input waveguides were adiabatically tapered out towards the edge of the chip to increase the spacing between them to be 10 µm which was ~ 10× the size of the focused laser spot used for coupling the input light. This ensured that only one input waveguide was excited at a time during the measurement of the transmission matrix.

Experimental measurement: To measure the transmission matrix of our encoder (Figure 5), we used an aspheric lens to couple continuous wave (CW) laser light at $\lambda = 1550$ nm to the chip. The lens was used to focus the laser beam to a spot of diameter ~1.5 µm at the edge of the waveguides. To measure the transmission matrix, the focused laser spot was scanned to couple to each input waveguide while the transmitted speckle pattern was recorded from above using a long-working distance objective (50x, NA=0.7) and an InGaAs camera (Xenics, Cheetah). All the images acquired during the measurement were then processed to determine the transmission matrix which we used for encoding the images. To quantify the experimental noise, the measurements were repeated and the difference in magnitude of the elements of the transmission matrix was used as the experimental noise present in encoding the images.

**Figures**

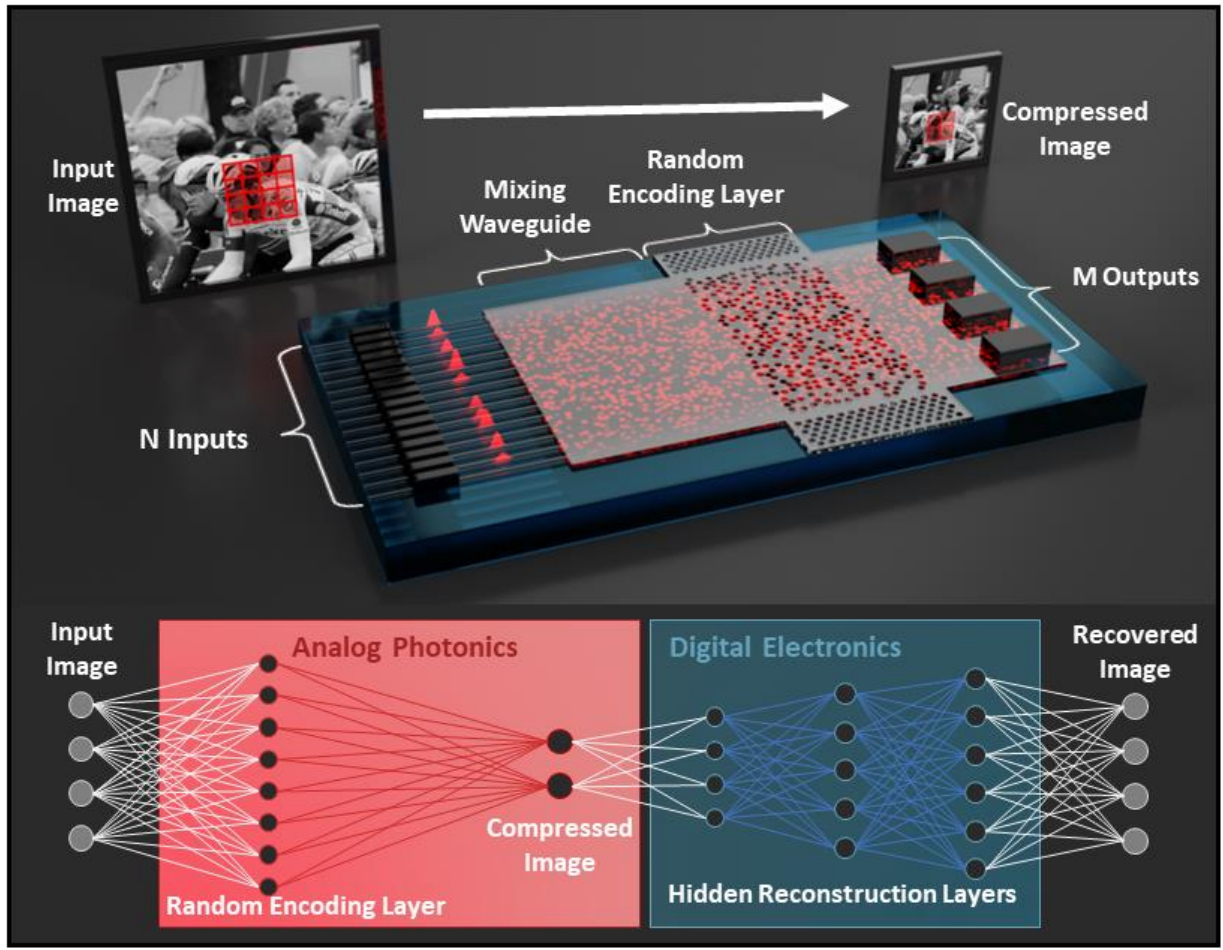

**Figure 1: Working principle of the photonic image encoder.** The silicon photonics-based all-optical image encoder consists of a series of *N* single mode input waveguides which carry information of the pixel of the images in the optical domain representing a $\sqrt{N} \times \sqrt{N}$ pixel block of the input image. The input waveguides connect to a multimode silicon waveguide region followed by a disordered scattering region which encodes the input through a local random transformation for image compression. The encoded output, which is a random looking spatially varying intensity pattern, is binned into *M* non-overlapping spatial regions (corresponding to *M* detectors) such that $M < N$ and the entire image is compressed through a series of these block-wise transforms. While the compression is performed optically, the reconstruction and image conditioning steps are performed electronically at the backend and the entire compression and reconstruction scheme is shown above.

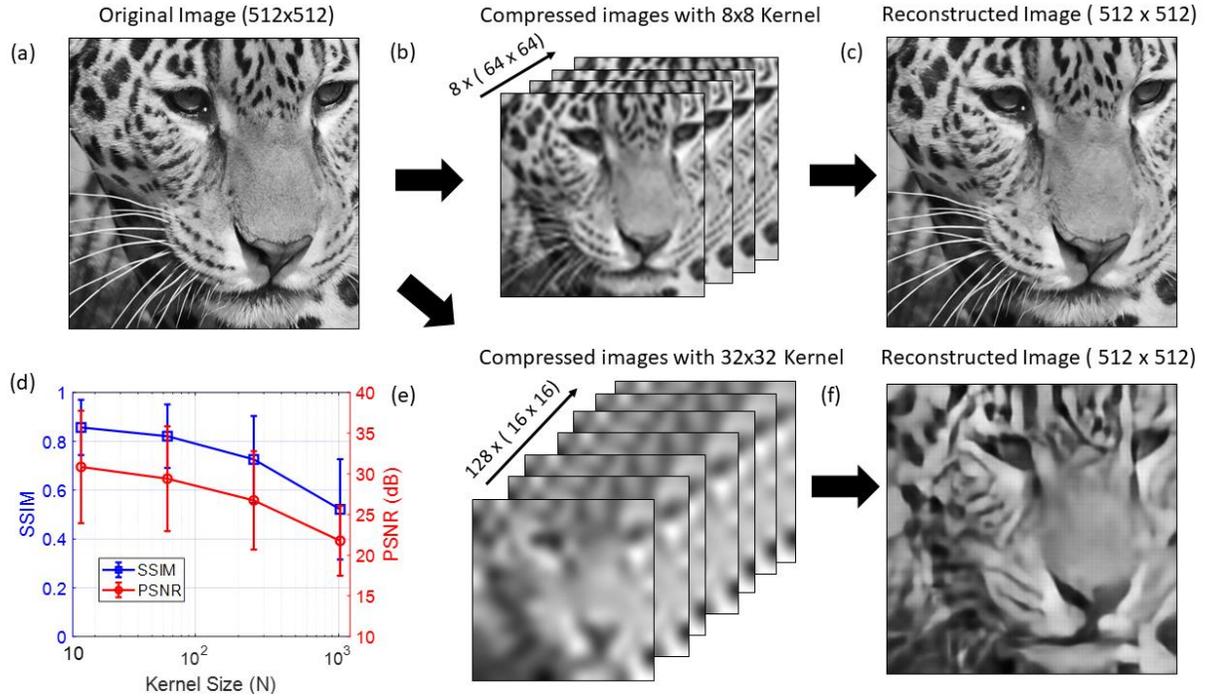

**Figure 2: Effect of kernel size on image compression and reconstruction. (a)** One of the original (512 x 512 pixel) grayscale noiseless images used to demonstrate compression. **(b, e)** Compressed images using an 8x8 (b) and 32 x 32-pixel kernel (e). For both cases, the compression ratio ($N:M$) is fixed to be 8:1 where in (b) $N = 64$, and $M = 8$ and in (e) $N = 1024$, and $M = 128$. The sizes of the compressed images are shown in brackets **(c, f)** Reconstructed images (512x512-pixel) from the compressed images shown in (b) and (e) respectively. **(d)** Mean PSNR and SSIM of the reconstructed images from the test data as a function of the kernel size. The error bars correspond to 1 standard deviation.

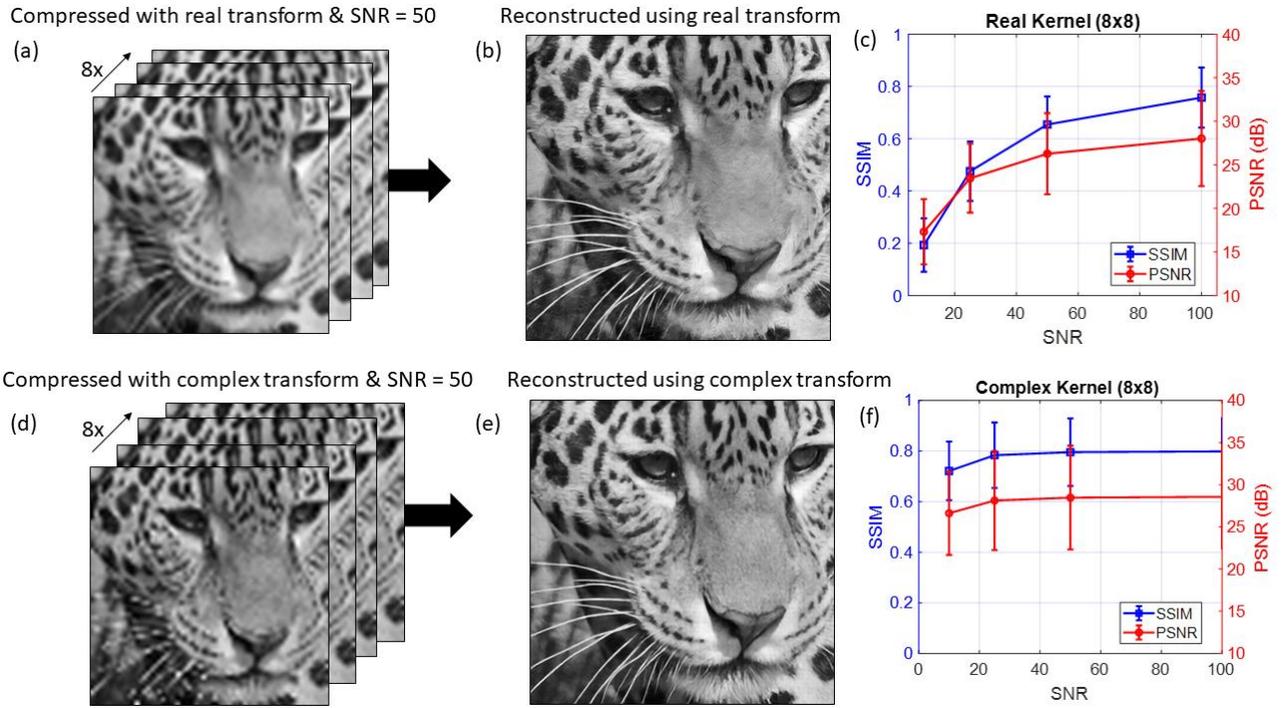

**Figure 3: Effect of kernel type (real vs complex) on image compression and reconstruction.** (**a**) 64x64-pixel images that were compressed using an 8x8 real kernel with 2 % gaussian white noise (SNR = 50). The original image was 512 x 512 pixels and is shown in Fig. 2(a). (**b**) Reconstructed image (512x512-pixel) from the compressed images shown in (a). (**c**) Mean PSNR and SSIM of the reconstructed images from the test data as a function of SNR. The error bars correspond to 1 standard deviation. (**d**) 64x64-pixel images that were compressed using an 8x8 complex kernel with 2 % gaussian white noise (SNR = 50). The original image was 512 x 512 pixels and is shown in Fig. 2(a). (**e**) Reconstructed image (512x512-pixel) from the compressed images shown in (d). (**f**) Mean PSNR and SSIM of the reconstructed images from the test data as a function SNR. The error bars correspond to 1 standard deviation.

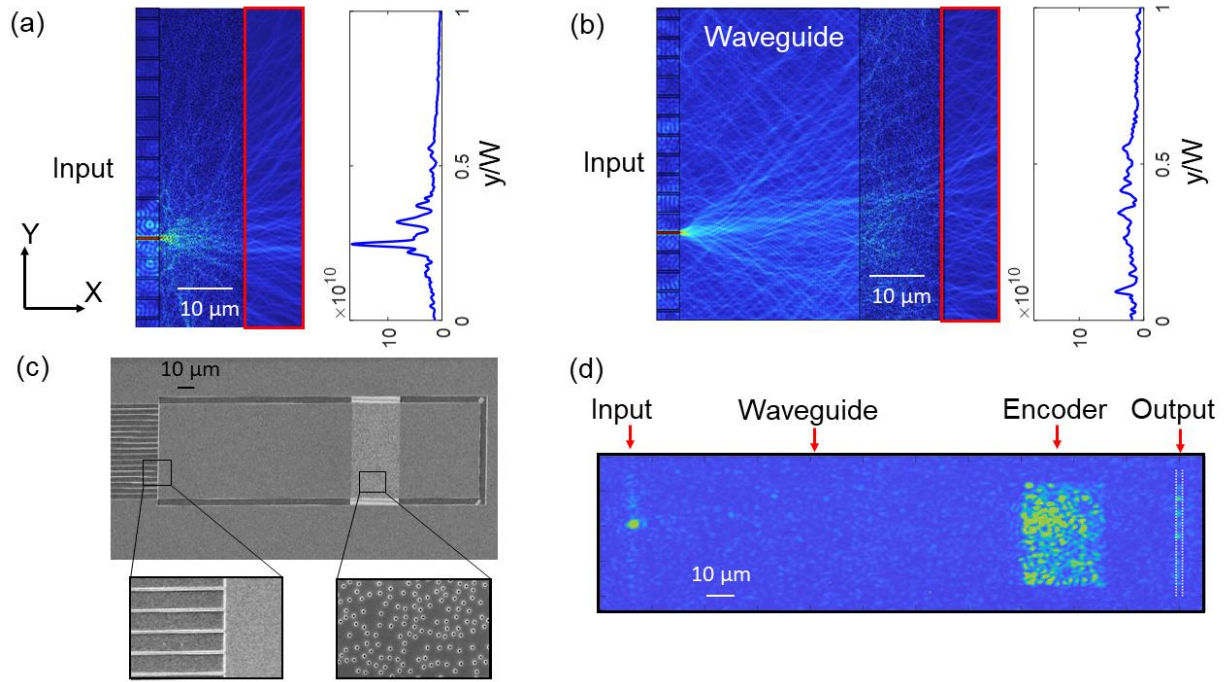

**Figure 4: Numerical simulations and experimental characterization. (a, b)** Full-wave frequency domain numerical simulations of the optical encoder with no (a) and a 32 µm multimode waveguide (b) in front of a 15 µm scattering region. Addition of the multimode waveguide region leads to spreading of the input light laterally along Y resulting in a random transmission matrix. **(c)** Scanning electron micrograph of the fabricated silicon-photonics based all-optical image encoder. **(d)** An example of a typical experimental measurement performed for characterization of the transmission matrix of the encoder.

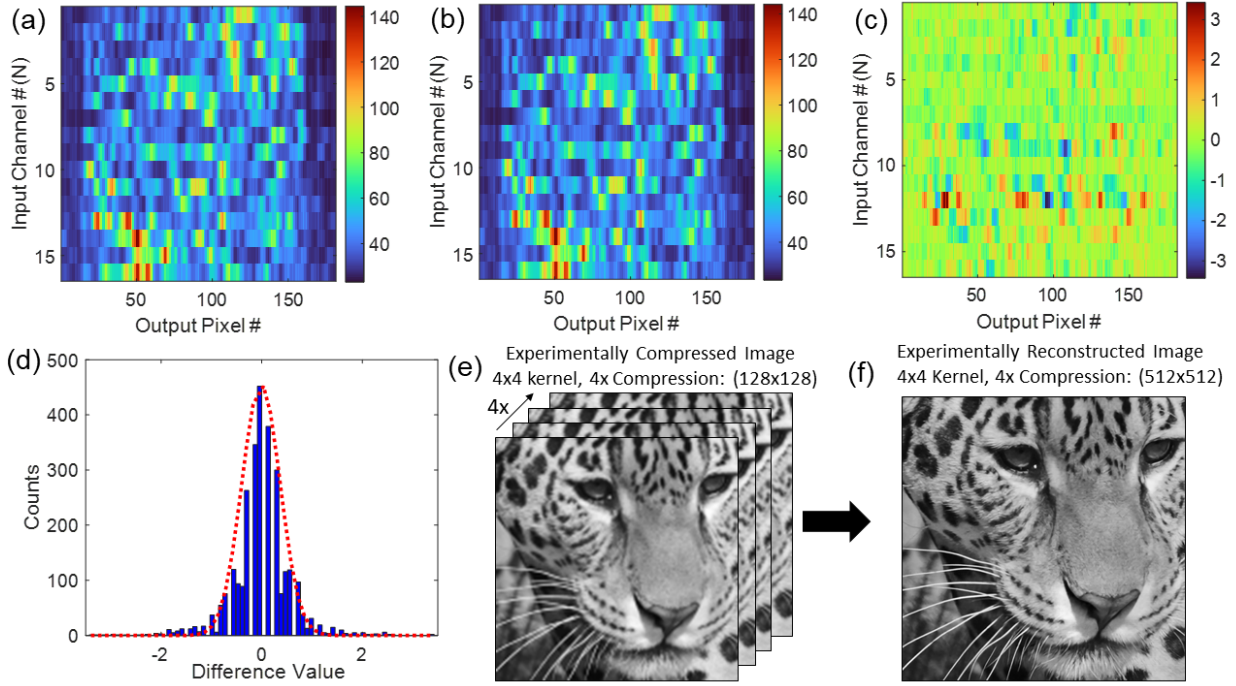

**Figure 5: Experimental demonstration of denoising and image compression. (a, b)** Experimental measurements of transmission matrices of the encoder shown in Fig. 4(c). The two measurements shown in (a) and (b) are of the same device but measured at different time intervals. The transmission matrices are extracted from a series of measurements similar to the one shown in Fig. 4(d). The x-axis of the data corresponds to number of pixels at the output (image shown in Fig. 4(d)) and y-axis corresponds to the number of the input channels. **(c)** 2D plot showing the difference in magnitude of the different elements of the transmission matrices measured in (a) and (b). The difference in magnitude is on the order of ~ 1 %. **(d)** Histogram showing the distribution of the difference values measured in (c). The distribution corresponds to a gaussian distribution centered around 0. The red-dotted line shows the gaussian function fitted to the experimentally measured data. The red-dotted gaussian function was used as the noise source in the reconstruction algorithms. **(e)** An example of one of many compressed images where the compression was done using the experimentally measured transmission matrix. The original image is shown in Fig. 2(a) and compression ratio is 4:1 with each image being a 128x128 pixel image. **(f)** Reconstructed 512x512-pixel image from the compressed image shown in (e).

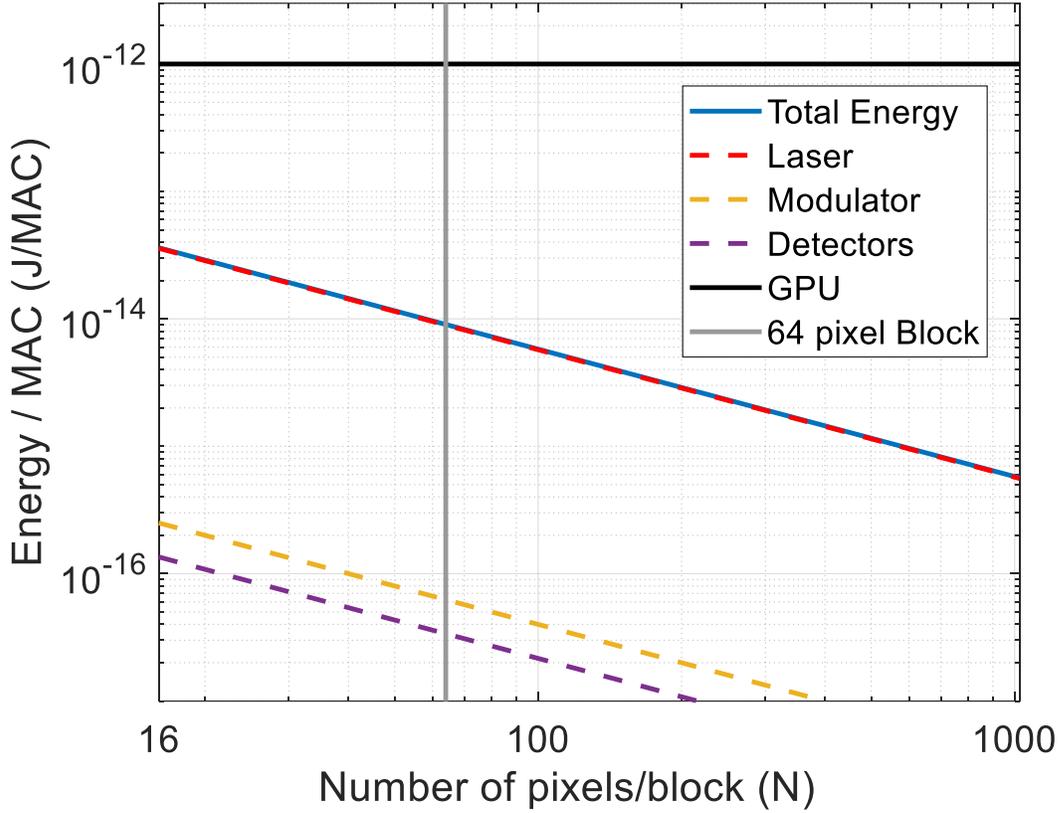

**Figure 6: Comparison of energy consumption for electronic and all-optical encoding approaches for image compression.** Comparison of energy consumption (given by Energy/MAC) as a function of number of inputs pixels *N* for encoding via matrix multiplication using a GPU-based electronic approach (black solid line) and the presented optical approach using a photonic encoder with average transmission $T_{scatter} = 20$ % (blue solid line). Different color-coded lines correspond to the different components contributing to the total energy consumption such as the laser (dashed red line), modulators (dashed yellow lines), and detectors (dashed magenta line). The vertical gray line corresponds to energy consumption values for an 8x8 kernel or a 64-input pixel block. The energy consumption is dominated by the input laser and the overall energy efficiency of the presented optical approach improves as N or number of input pixels increases.

# Integrated Photonic Encoder for Terapixel Image Processing

Xiao Wang [1, +], Brandon Redding [2, +], Nicholas Karl [3], Christopher Long [3], Sean Pang [4], David Brady[1], Raktim Sarma[3, 5, *]

[+]Equal contribution

[*] rsarma@sandia.gov

[1] Wyant College of Optical Sciences, University of Arizona, Tucson, Arizona, USA

[2] U.S. Naval Research Laboratory, Washington, DC, USA

[3]Sandia National Laboratories, Albuquerque, New Mexico, USA

[4]CREOL, The College of Optics and Photonics, University of Central Floria, Orlando, Florida, USA

[5]Center for Integrated Nanotechnologies, Sandia National Laboratories, Albuquerque, New Mexico, USA

## S1. Additional Experimental results: Compressed and reconstructed images and their statistics.

In the main text, we presented only one example of experimentally compressed and reconstructed image. In Fig S1, we present some more examples of experimentally compressed and reconstructed images from the test dataset. The first row in Fig. S1 corresponds to the original images. The second and third row show the experimentally compressed and reconstructed images. The sizes of the images are shown in brackets. For the compressed images, only one of the four compressed images are shown here. In Fig. S2 we show the histogram of PSNR and SSIM of all the reconstructed images from the test dataset that we compressed using our experimentally measured encoding matrix.

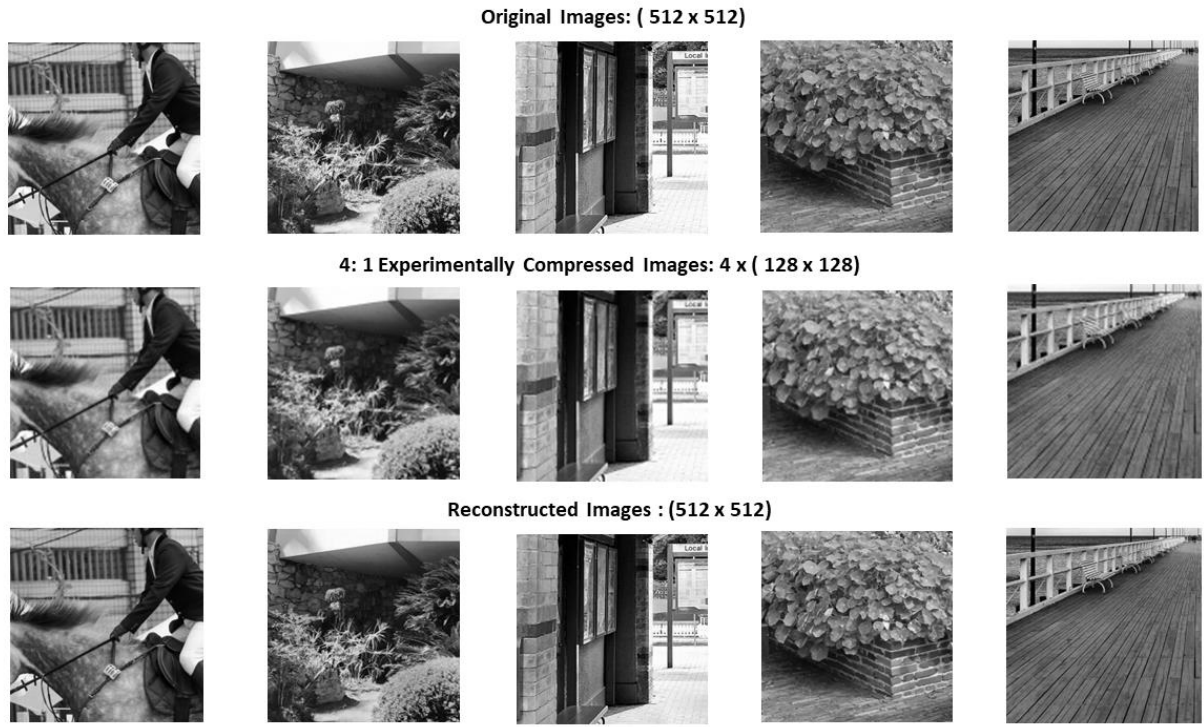

**Figure S1: Examples of experimentally compressed images.** The first row corresponds to original images taken from the test dataset. The second row are corresponding compressed images that are compressed using the experimentally measured encoding matrix. Only one of the 4 compressed images are shown. The third row corresponds to the reconstructed images. The sizes of the images are shown in brackets.

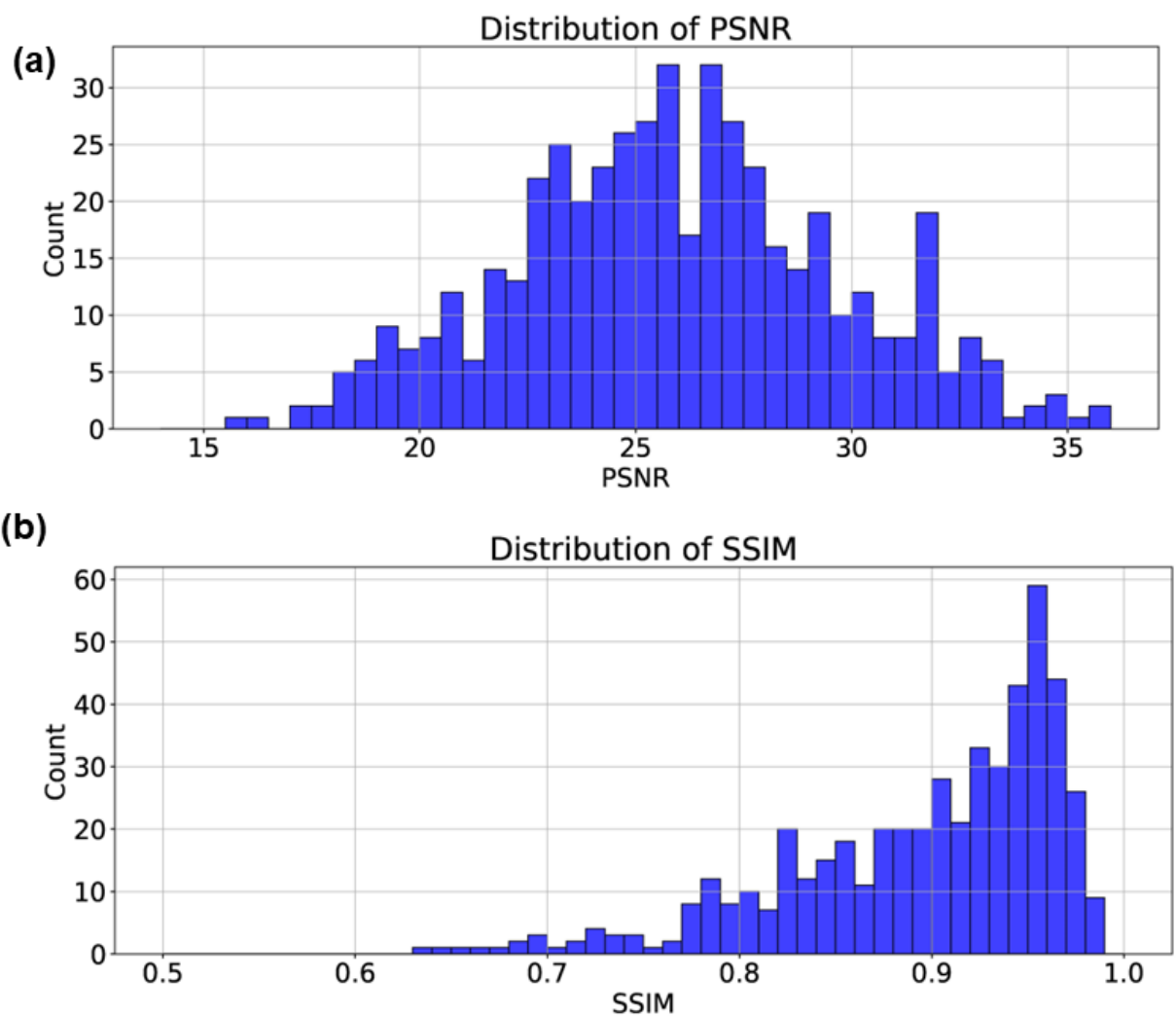

**Figure S2: Statistics of reconstructed images that were compressed using experimentally measured encoding matrix.** (**a**) Histogram of PNSR all the reconstructed images from the test dataset. (**b**) Histogram of SSIM all the reconstructed images from the test dataset.

## S2. Energy consumption typical of mainstream electronic architectures

Today's digital accelerator landscape includes various parallel chip architectures with a range of core counts and performance metrics. Table 1 summarizes mainstream architectures and the energy consumption for these major categories.

| Type | Name | Energy Efficiency |
|------|------|-------------------|
| GPU | NVIDIA V100 | 4.6 pJ/MAC |
|  | NVIDIA H100 SXM | 0.7 pJ/MAC |
| ASIC | Google TPU v1 | 0.8 pJ/MAC |
|  | Google TPU v4 | 1.2 pJ/MAC |
| FPGA | Xilinx FPGA Alveo U250 | 13.5 pJ/MAC |
|  | Xilinx FPGA Versal VC2802 | 0.7 pJ/MAC |

**Table 1:** Energy efficiency of prominent categories of digital accelerators. Typical energy efficiency of these digital accelerators is on the order of ~ 1 pJ/MAC.

## S3. Integration of photonic encoder with silicon photonics and CMOS components

In the main text, we presented a passive, proof-of-concept device and showed experimental compression using a real-valued transmission matrix. While the passive photonic encoder forms the most critical component of the image processing engine, the complete optoelectronic image processing engine will require integration of active CMOS components such as modulators and detectors which will also enable operation using a complex-valued transmission matrix. This is because integration of modulators and detectors will allow us to extract both amplitude and phase of the encoded electric fields for different inputs using phase retrieval compressive measurement techniques [ref? maybe a phase retrieval paper if David has one? Not crucial though]. As shown in the main text, compression of images using complex transforms is more robust to experimental noise.

A high-level view of the photonic accelerator and electronic interface of our envisioned image processing engine is shown in Fig. S3. The optical components are inside the dotted box, and the CMOS electronic components are external to it. The optoelectronic accelerator is designed to convert $N$ inputs into $M$ outputs, where $N$ and $M$ can be different depending on the computing task. To convert data to the optical domain, a set of $N$ modulators will be driven by analog signals that are transduced onto an optical carrier provided by a laser. Light then propagates through the accelerator device and is received by $M$ photodetectors. The optical operation will be supported by multiple CMOS components: the driver sets the output voltage levels for the modulator, the transimpedance amplifier (TIA) converts the small current signal into a voltage and a secondary amplifier sets the voltage level for the CMOS family. At either end of the link are the digital/analog (D/A) and analog/digital (A/D) converters.

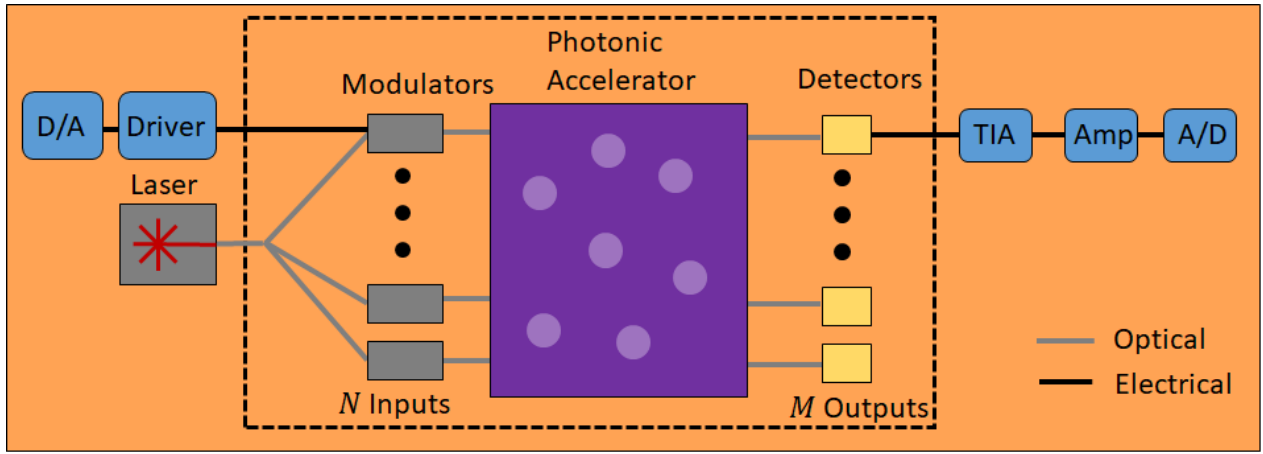

**Figure S3.** High-level view of the optoelectronic image processing engine. The complete image processing engine will include the photonic accelerator, modulators, detectors, and CMOS ICs. The silicon photonic components and CMOS components are placed inside and outside the dotted box, respectively. In general, a single laser (assumed to be off-chip here) could be used to drive multiple photonic accelerators.